\documentclass[pra,superscriptaddress,floatfix,amsmath,footinbib,amssymb,twocolumn]{revtex4-2}
\usepackage[utf8]{inputenc}
\usepackage[english]{babel}
\usepackage[T1]{fontenc}
\usepackage{graphicx}
\usepackage{dcolumn}
\usepackage{bm}
\usepackage{booktabs}
\usepackage{array}
\usepackage{hhline}
\usepackage{physics}
\usepackage{mathrsfs}
\usepackage{calligra}
\usepackage{dsfont}
\usepackage[table]{xcolor}
\usepackage{colortbl}
\usepackage{caption}
\usepackage{subcaption}
\usepackage{ragged2e}
\usepackage{soul}
\usepackage[x11names]{xcolor}
\usepackage{hyperref}
\hypersetup{
    colorlinks=true,
    linkcolor=green,
    filecolor=magenta,      
    urlcolor=blue,
}

\DeclareMathOperator{\sinc}{sinc}
\DeclareMathAlphabet{\mathpzc}{OT1}{pzc}{m}{it}

\begin{document}


\title{Gravitational wave imprints on spontaneous emission}

\author{Jerzy Paczos}
\email{jerzy.paczos@fysik.su.se}
\affiliation{Department of Physics, Stockholm University, SE-106 91 Stockholm, Sweden}

\author{Navdeep Arya}
\email{navdeep.arya@fysik.su.se}
\affiliation{Department of Physics, Stockholm University, SE-106 91 Stockholm, Sweden}

\author{Sofia Qvarfort}
\email{sofia.qvarfort@fysik.su.se}
\affiliation{Department of Physics, Stockholm University, SE-106 91 Stockholm, Sweden}
\affiliation{Nordita, KTH Royal Institute of Technology and Stockholm University, Hannes Alfv{\'e}ns v{\"a}g 12, SE-114 19 Stockholm, Sweden}

\author{Daniel Braun}
\email{daniel.braun@uni-tuebingen.de}
\affiliation{Institut für Theoretische Physik, Eberhard Karls Universität Tübingen, 72076 Tübingen, Germany}

\author{Magdalena Zych}
\email{magdalena.zych@fysik.su.se}
\affiliation{Department of Physics, Stockholm University, SE-106 91 Stockholm, Sweden}

\begin{abstract}
    Despite growing interest, there is a scarcity of known predictions in the regime where both quantum and general relativistic effects become observable. Here, we investigate a combined atom-field system in a curved spacetime, with a specific focus on gravitational-wave backgrounds. We demonstrate that a plane gravitational wave alters spontaneous emission from a single atom, manifesting itself as a direction-dependent change in the emission spectrum. Although the total decay rate remains unchanged, implying that no information about the gravitational wave is stored in the atomic internal state alone, the wave leaves imprints on the evolution of the composite atom-field system. To quantify how well this effect can be measured, we analyze both the classical Fisher information associated with photon number measurements and the quantum Fisher information. Our analysis indicates that the effect could be measured in state-of-the-art cold-atom experiments and points to spontaneous emission as a potential probe of low-frequency gravitational waves.
\end{abstract}

\maketitle

\textit{Introduction---}Currently testable general relativistic effects in Earth-based experiments comprise gravitational redshift tests~\cite{Bothwell2022, Zheng2022} and gravitational wave (GW) detection~\cite{Acernese2015, Aasi2015}. In this context, GWs stand out as their measurements directly support the dynamical, curved spacetime predicted by general relativity. Since Einstein predicted GWs in 1916~\cite{Einstein1916} as ripples in the fabric of spacetime propagating at the speed of light, theoretical and experimental efforts have largely concentrated on understanding how these waves affect the motion of test masses~\cite{Pirani1956, Bondi1959, Weber1960, Weber1966, Weber1967, Weber1968}. Early analyses focused on the geodesic deviation induced by passing waves, which is at the heart of modern detection techniques and proposals, including ground-based laser interferometers~\cite{Acernese2015, Aasi2015}, resonant bar detectors~\cite{Weber1960, Weber1966, Weber1967, Weber1968}, pulsar timing arrays~\cite{Sazhin1978, Detweiler1979, Hellings1983}, space-based observatories~\cite{Armstrong1999}, atom interferometers~\cite{Dimopoulos2008}, levitated masses~\cite{Arvanitaki2013}, and optical atomic clocks~\cite{Kolkowitz2016}. Despite their diversity, these approaches are unified by a common framework: the detection of GWs through their classical influence on the trajectories or separations of test masses.

Identifying testable general relativistic effects by means beyond classical test masses has been of increasing interest and importance in the foundations of physics. In this context, notable early steps towards exploring the effects of GWs involved studying their direct interaction with quantum systems. The interaction with a hydrogen atom was shown to induce small changes in its internal energy levels~\cite{Leen1983} and resonant quadrupole transitions~\cite{Fischer1994}. These effects arise from the finite spatial extent of the atom and its quadrupole coupling to the GW amplitude and are potentially relevant for highly excited Rydberg atoms. The effect of GWs on other quantum systems has been studied in the context of particle production~\cite{Gibbons1975, Deser1975, Sorge2000, Sabin2014, Jones2017, Gray2021} and scattering~\cite{Garriga1991}, Rabi oscillations~\cite{Siparov2004, Sorge2023}, field-mediated entanglement~\cite{Xu2020, Barman2023, Barman2025}, and GW-induced superradiance in atomic arrays~\cite{Arya2024}.

In this Letter, we investigate how a plane GW background affects spontaneous emission from a point-like two-level atom. We show that a GW-induced periodic modulation of the quantum field interacting with the atom~\cite{Garriga1991} gives rise to an angle-dependent correction to the emission spectrum. In the limit of large GW frequency, this leads to sidebands in the emission spectrum --- a hallmark of periodically driven systems~\cite{Autler1955, Mollow1969, Grove1977, Boca2004, Maunz2005}. On the other hand, in the low GW frequency regime, the effect reduces to an angle-dependent frequency shift of the emitted photons. Moreover, the GW-induced correction exhibits a characteristic quadrupolar shape in the plane perpendicular to the GW propagation --- a feature that may help distinguish this effect from other factors affecting the spectrum. Interestingly, the total emission rate remains unchanged, indicating that no information about the GW is imprinted on the internal state of the atom. It is the quantum field --- an extended system --- that plays the central role by encoding information about the GW.

To assess the extractable information, we consider the estimation of the GW amplitude and analyze both the classical Fisher information associated with photon number measurements and the quantum Fisher information encoded in the atom-field state. We show that for a given GW frequency, there are optimal system-evolution times for which the quantum Cramér–Rao bound~\cite{Braunstein1994, Braunstein1996} can be saturated using photon number measurements. This is analogous to the optimal effective length in interferometric detectors~\cite{Maggiore2007}. Our analysis falls within the general framework for estimating spacetime parameters based on quantum measurements~\cite{Downes2017}. However, rather than focusing merely on the fields and working in the large-mean-field approximation (as in~\cite{Downes2017}), we consider the atom-field interaction in the weak-field regime (vacuum state), and our results do not follow from prior works. Finally, we estimate the minimal number of atoms required for GW detection based on the analysis of the classical and quantum Fisher information, and find that the requirements are not daunting. In particular, for shot-noise limited experiments, to reach the typically desired sensitivity~\cite{Robson2019, Babak2021} in the (sub-)millihertz range, one would need at least $10^6-10^8$ atoms (numbers which have already been achieved in experiments with cold atomic clouds~\cite{Labeyrie1999, Guerin2016, Kashanian2016}).

The method developed in this work can be applied to any spacetime metric on which quantum field theory has been developed~\cite{Birrell1982, Wald1994, Mukhanov2007}. This opens up the possibility of studying joint general relativistic and quantum effects in other spacetime geometries based on the effects in spontaneous emission proposed here. Throughout, we use natural units $\hbar=c=1$.

\textit{Setup---}Consider a simplified model of atom-light interaction, where the atom is treated as a point-like two-level system with the ground state $\ket{g}$, the excited state $\ket{e}$, and the energy gap $\omega_0$, and the light is modeled as a massless real scalar field $\hat{\phi}$. The field is assumed to couple linearly to the monopole moment of the atom $\hat{m}(\tau)$
through the (interaction picture) Hamiltonian
\begin{equation}\label{Hamiltonian}
    \hat{H}_\text{I}(\tau)=\varepsilon\hat{m}(\tau)\hat{\phi}(x(\tau)),
\end{equation}
where $\tau$ is the proper time of the atom, $\varepsilon\ll1$ is a small coupling constant, and $x(\tau)=(t(\tau),\boldsymbol{x}(\tau))$ is the atomic trajectory. The monopole moment can be expressed in terms of raising and lowering operators $\hat{\sigma}_+\equiv\ket{e}\bra{g}$ and $\hat{\sigma}_-\equiv\ket{g}\bra{e}$ as $\hat{m}(\tau)=\mathrm{e}^{\mathrm{i}\omega_0 \tau}\hat{\sigma}_+ + \mathrm{e}^{-\mathrm{i}\omega_0 \tau}\hat{\sigma}_-$. Given the spacetime metric and the complete set $\{u_{\boldsymbol{k}},u_{\boldsymbol{k}}^*\}_{\boldsymbol{k}}$ of orthonormal solutions (modes) to the Klein-Gordon equation for the scalar field on that spacetime, we expand the field operator in the standard way:
\begin{equation}\label{field operator}
    \hat{\phi}(x(\tau))=\int\dd^3\boldsymbol{k}\left[u_{\boldsymbol{k}}(x(\tau))\hat{a}_{\boldsymbol{k}}+u_{\boldsymbol{k}}^*(x(\tau))\hat{a}_{\boldsymbol{k}}^\dagger\right],
\end{equation}
where the annihilation and creation operators $\hat{a}_{\boldsymbol{k}}$ and $\hat{a}_{\boldsymbol{k}}^\dagger$ satisfy $[\hat{a}_{\boldsymbol{k}},\hat{a}_{\boldsymbol{l}}^\dagger]=\delta^{(3)}(\boldsymbol{k}-\boldsymbol{l})$ with other commutators vanishing. Information about the metric in the field expansion~\eqref{field operator} is encoded in the modes $u_{\boldsymbol{k}}$.

The assumption of point-like atoms and the use of the interaction Hamiltonian~\eqref{Hamiltonian} exclude any direct coupling between the atoms and gravity. As a result, the small internal energy shifts~\cite{Leen1983} and resonant atom-GW interactions~\cite{Fischer1994} (both potentially relevant for highly excited Rydberg atoms) are neglected in the present analysis. The effects described in this work result solely from the change of the field modes $u_{\boldsymbol{k}}$ due to gravity and are independent of the atom size (as long as it is much smaller than the GW wavelength and the atomic transition wavelength so that the point-like approximation is valid).

The model constructed in Eq.~\eqref{Hamiltonian} is known to capture the main features of the atom-light interaction under the assumption of interactions without exchange of angular momentum~\cite{Martinez2013, Alhambra2014}. Note that GW effects on the total transition rate have been studied using this model, e.g., in~\cite{Chen2022, Prokopec2023}, showing only quadratic dependence on the GW amplitude. In contrast, we show that the emission spectrum and directionality are affected in the first order in the GW amplitude and are, therefore, more relevant for potential experiments.

\textit{Spontaneous emission---}We assume the initial atom-field state (at time $\tau_i$) of the form $\ket{\psi_0}=\ket{e}\otimes\ket{0}$, where $\ket{0}$ is the field vacuum state defined by $\hat{a}_{\boldsymbol{k}}\ket{0}=0$, and consider the state of the system at a later time $\tau_f$:
\begin{equation}
    \ket{\psi(\tau_f,\tau_i)}=\alpha(\tau_f,\tau_i)\ket{e,0}+\int\mathrm{d}^3\boldsymbol{k}\beta_{\boldsymbol{k}}(\tau_f,\tau_i)\ket{g,1_{\boldsymbol{k}}},
\end{equation}
where $\ket{1_{\boldsymbol{k}}}\equiv\hat{a}_{\boldsymbol{k}}^\dagger\ket{0}$ is a single-photon state with the photon in mode $\boldsymbol{k}$. Up to the second order in $\varepsilon$, the coefficients $\alpha$ and $\beta_{\boldsymbol{k}}$ are given by
\begin{equation}\label{state coefficients}
    \begin{split}
        &\alpha(\tau_f,\tau_i)=1-\varepsilon^2\int_{\tau_i}^{\tau_f}\mathrm{d}\tau\int_{\tau_i}^{\tau}\mathrm{d}\tau'\mathrm{e}^{\mathrm{i}\omega_0(\tau-\tau')}\times\\
        &\hspace{2.25cm}\times\int\mathrm{d}^3\boldsymbol{k}u_{\boldsymbol{k}}(x(\tau))u_{\boldsymbol{k}}^*(x(\tau')),\\
        &\beta_{\boldsymbol{k}}(\tau_f,\tau_i)=-\mathrm{i}\varepsilon\int_{\tau_i}^{\tau_f}\mathrm{d}\tau\mathrm{e}^{-\mathrm{i}\omega_0\tau}u_{\boldsymbol{k}}^*(x(\tau)).
    \end{split}
\end{equation}
and to calculate them, we need to know the modes $u_{\boldsymbol{k}}$ and the trajectory $x(\tau)$. Eq.~\eqref{state coefficients} is valid for an arbitrary trajectory in a general spacetime in which quantum field theory is well-defined~\cite{Birrell1982, Wald1994, Mukhanov2007}, and therefore leads to testable predictions of quantum field theory in curved spacetimes.

As a specific example, consider the background metric corresponding to the plane, plus-polarized GW with amplitude $\mathcal{A}$ and frequency $\omega$ in the transverse-traceless gauge~\cite{Maggiore2007}
\begin{equation}\label{metric}
    \begin{split}
        \dd s^2=&-\dd t^2+(1+\mathcal{A}\cos[\omega(t-z)])\dd x^2\\&+(1-\mathcal{A}\cos[\omega(t-z)])\dd y^2+\dd z^2.
    \end{split}
\end{equation}
In this metric, trajectories with constant spatial coordinates (i.e., time independent) are geodesics, and $\tau=t$ is their proper time. Assume that the atom follows a geodesic with $x(t)=(t,\boldsymbol{0})$. The choice of $x$ and $y$ coordinates is arbitrary since the metric is homogeneous in the $xy$ plane, while the selection of the $z$ coordinate corresponds to the choice of the initial phase of the GW. With this choice of the atomic position, we can suppress the $\boldsymbol{x}$-dependence of the field operator by defining $\hat{\phi}(t)\equiv\hat{\phi}(t,\boldsymbol{0})$ and $u_{\boldsymbol{k}}(t)\equiv u_{\boldsymbol{k}}(t,\boldsymbol{0})$.

\begin{figure*}[t!]
    \begin{subfigure}[t]{0.48\textwidth}
        \centering
        \includegraphics[width=0.9\textwidth]{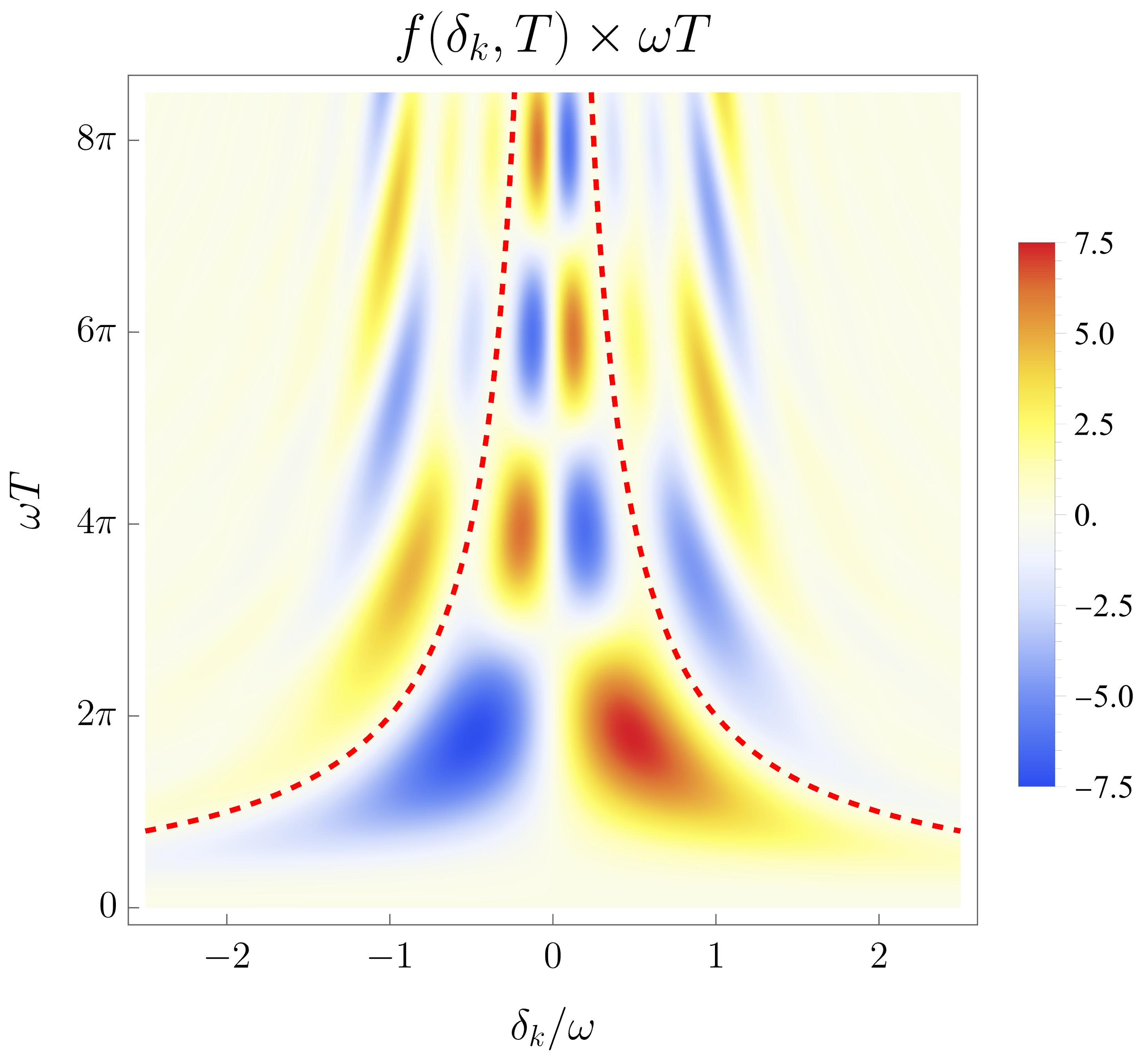}
        \caption{}
        \label{fig: f}
    \end{subfigure}%
    ~
    \centering
    \begin{subfigure}[t]{0.48\textwidth}
        \centering
        \includegraphics[width=0.9\textwidth]{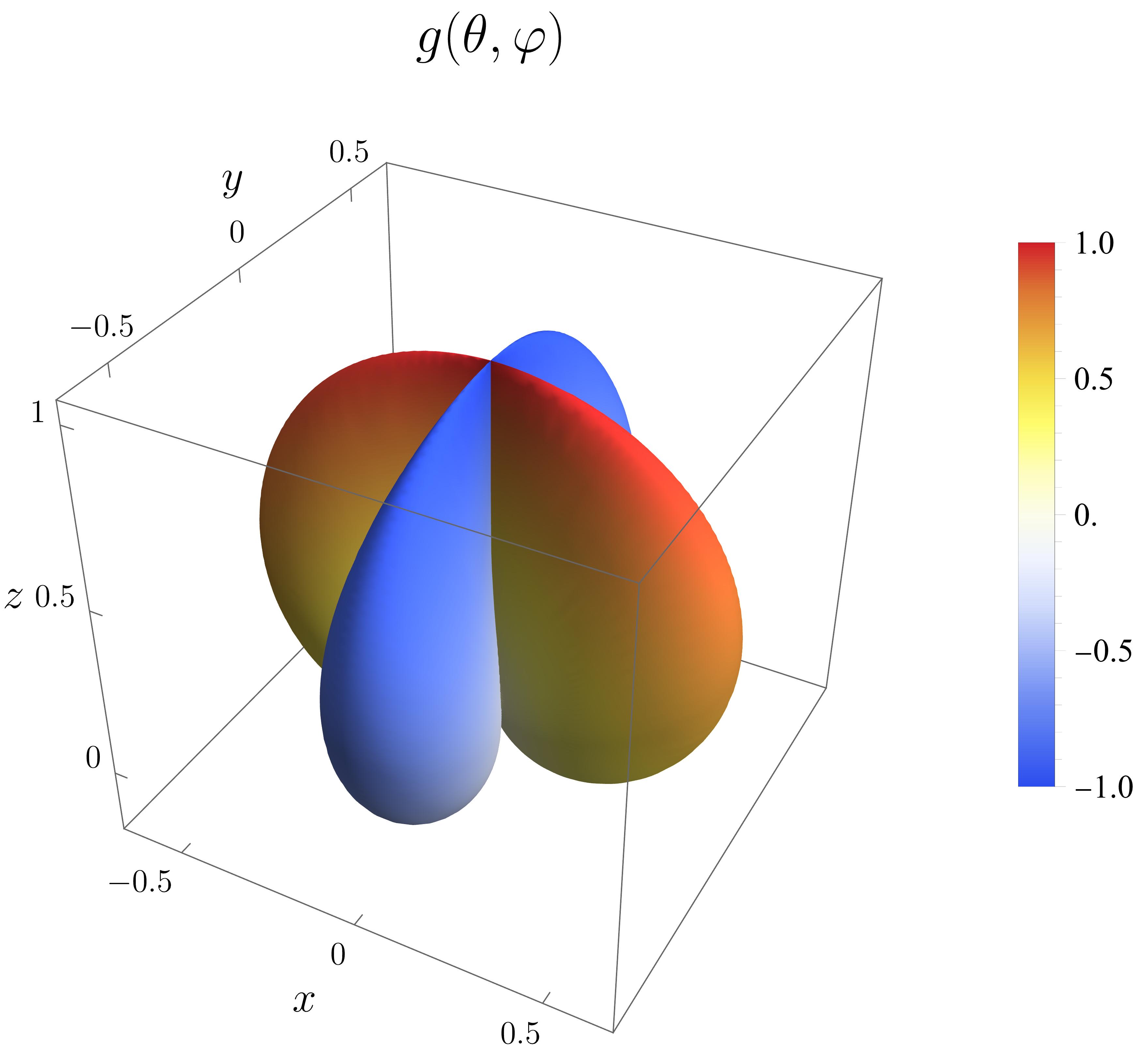}
        \caption{}
        \label{fig: g}
    \end{subfigure}
    \caption{\justifying Functions $f(\delta_k,T)$ and $g(\theta,\varphi)$ governing frequency and angular dependence of the GW correction to the photon emission. Their values are indicated by color, with blue and red corresponding to negative and positive values. From (a), it is apparent that the GW gives rise to positive and negative corrections on the sides of the carrier frequency ($\delta_k=0$). The corrections are antisymmetric with respect to the carrier frequency. The red dashed lines in the plot correspond to $\delta_kT=\pm2\pi$, designating the width of the flat spacetime contribution $\langle\tilde{n}_{\boldsymbol{k}}\rangle$. From the function $g(\theta, \varphi)$ in panel (b), it is apparent that the effect is maximal along the positive $z$ direction (direction of GW propagation) and vanishes in the opposite direction. It exhibits a quadrupolar pattern in the $xy$ plane with opposite signs along the $x$ and $y$ axes. The distance from the origin additionally highlights the absolute value of the function $g(\theta,\varphi)$ at given $\theta$ and $\varphi$.}
    \label{fig:g and f}
\end{figure*}

Given the metric~\eqref{metric}, we can write the corresponding Klein-Gordon equation and construct the orthonormal solutions (see~\cite{Garriga1991}) needed for the expansion~\eqref{field operator}. We introduce spherical coordinates in momentum space, $\boldsymbol{k}=k(\sin\theta\cos\varphi,\sin\theta\sin\varphi,\cos\theta)$, and a function $C_{\boldsymbol{k}}=\mathcal{A}\frac{k}{\omega}g(\theta,\varphi)$ with $g(\theta,\varphi)\equiv\cos^2(\theta/2)\cos(2\varphi)$. Then, the modes on the GW background are given by (see Appendix~\ref{Appendix: orthonormal modes})
\begin{equation}\label{mode solutions}
    u_{\boldsymbol{k}}(t)=\frac{1}{\sqrt{(2\pi)^32k}}\mathrm{e}^{-\mathrm{i}[kt-C_{\boldsymbol{k}}\sin(\omega t)]},
\end{equation}
corresponding to the flat-spacetime plane waves with periodically modulated phase.

For the metric~\eqref{metric} and the choice of mode solutions~\eqref{mode solutions} the coefficient $\beta_{\boldsymbol{k}}$ from Eq.~\eqref{state coefficients} is given by
\begin{equation}\label{amplitude sidebands}
    \begin{split}
        \beta_{\boldsymbol{k}}(t_f,t_i)\propto&\sum_{n=-\infty}^\infty J_n\left(C_{\boldsymbol{k}}\right)\mathrm{e}^{-\mathrm{i}n\left(\phi_i+\frac{\omega T}{2}\right)}\sinc\left[\frac{(\delta_k-n\omega)T}{2}\right],
    \end{split}
\end{equation}
where we suppressed the overall factor for brevity (see Appendix~\ref{Appendix: spontaneous emission} for details). In Eq.~\eqref{amplitude sidebands} we denote by $J_n\left(C_{\boldsymbol{k}}\right)$ the Bessel function of the first kind, $\phi_i\equiv\omega t_i$ is the initial phase of the GW, $\delta_k\equiv k-\omega_0$ is the atom-field-mode detuning, and $T\equiv t_f-t_i$ is the duration of the system evolution. Thus, the GW-induced modulation of the modes~\eqref{mode solutions} gives rise to infinitely many sidebands with frequencies shifted by $n\omega$, and amplitudes $\propto J_n\left(C_{\boldsymbol{k}}\right)$.

Given the coefficient $\beta_{\boldsymbol{k}}$, we can calculate the expected number of photons emitted into a specific mode, $\langle n_{\boldsymbol{k}}\rangle=|\beta_{\boldsymbol{k}}|^2$. The latter contains diagonal terms $\propto J_n^2\left(C_{\boldsymbol{k}}\right)$ corresponding to the individual spectral components, and cross-terms $\propto J_n\left(C_{\boldsymbol{k}}\right)J_m\left(C_{\boldsymbol{k}}\right)$ with $n\neq m$ representing interference between them. Up to the first order in $C_{\boldsymbol{k}}$ (and thus $\mathcal{A}$) the only terms that contribute are: the diagonal term $\propto J_0^2\left(C_{\boldsymbol{k}}\right)\approx 1$ (the carrier), and the interference terms between the carrier and the first sidebands $\propto J_0\left(C_{\boldsymbol{k}}\right)J_{\pm1}\left(C_{\boldsymbol{k}}\right)\approx\pm C_{\boldsymbol{k}}/2$. We refer to the first one, being identical to the result in flat spacetime, as the flat-spacetime contribution, and denote it by $\langle\tilde{n}_{\boldsymbol{k}}\rangle$. The latter is subsequently called the GW correction and denoted by $\langle\delta n_{\boldsymbol{k}}\rangle$. They are given by (see Appendix~\ref{Appendix: spontaneous emission})
\begin{equation}\label{emission contributions}
    \begin{split}
        &\langle \tilde{n}_{\boldsymbol{k}}\rangle=\frac{\gamma_0T^2}{8\pi^2\omega_0k}\sinc^2(\delta_k T/2),\\
        &\langle \delta n_{\boldsymbol{k}}\rangle=\frac{\gamma_0T^2}{8\pi^2\omega_0k}C_{\boldsymbol{k}}\cos(\phi_i+\omega T/2)f(\delta_k,T),
    \end{split}
\end{equation}
where $\gamma_0\equiv\varepsilon^2\omega_0/(2\pi)$ is the natural linewidth of the atomic transition, and $f(\delta_k,T)\equiv\sinc(\delta_k T/2)(\sinc[(\delta_k-\omega)T/2]-\sinc[(\delta_k+\omega)T/2])$.

The function $f(\delta_k,T)$ determines the frequency dependence of the GW-induced correction $\langle \delta n_{\boldsymbol{k}}\rangle$. It is plotted in Fig.~\ref{fig: f}, which shows that GW induces positive and negative corrections on the sides of the carrier frequency ($\delta_k=0$). The red dashed line in Fig.~\ref{fig: f} marks the points corresponding to $\delta_k T=\pm2\pi$, designating the width of the flat spacetime contribution $\langle\tilde{n}_{\boldsymbol{k}}\rangle$. For $\omega T\ll2\pi$, the function reduces to two peaks, one on each side of $\delta_k=0$ and both within the carrier range. This represents a small frequency shift of the carrier, $\Delta k/k=-\mathcal{A}\cos\phi_ig(\theta,\varphi)$, (see Appendix~\ref{Appendix: low frequency regime}). On the other hand, for $\omega T\gtrsim2\pi$ the correction $\langle\delta n_{\boldsymbol{k}}\rangle$ develops side peaks outside of the carrier (i.e., for $|\delta_k|>2\pi/T$).

The angular dependence of the correction $\langle \delta n_{\boldsymbol{k}}\rangle$ is dictated by the factor $C_{\boldsymbol{k}}\propto g(\theta,\varphi)$ (see Fig.~\ref{fig: g}). It has a characteristic quadrupolar shape in the $xy$ plane, reflecting the GW-induced deformation of space. Note that the correction is maximal along the direction of GW propagation and vanishes in the opposite direction. This directionality of the GW correction encodes information about the GW direction of propagation and its polarization, and is to be contrasted with the isotropy of the flat spacetime contribution.

Interestingly, the GW correction vanishes when integrated over the emitted photon's momentum, indicating that there is no net change in the total emission rate. Consequently, up to the first order in GW amplitude, no information about the GW is encoded in the atom’s internal state --- the atom acts only as a transducer of the information stored in the field.

\begin{table}[t!]
\renewcommand{\arraystretch}{1.5}
\setlength{\tabcolsep}{12pt}
\begingroup
\centering
\begin{tabular}{|l|c|c|}
\hhline{|---|}
\textbf{Source} & \textbf{Frequency} & \textbf{Amplitude} \\
\hhline{|---|}
MBHMs & $\sim 10^{-4}\text{--}10^{-1}\,\mathrm{Hz}$ & $\lesssim10^{-16}$ \\
\hhline{|---|}
EMRIs & $\sim 10^{-3}\text{--}10^{-1}\,\mathrm{Hz}$ & $\lesssim 10^{-18}$ \\
\hhline{|---|}
GBs & $\sim 10^{-4}\text{--}10^{-2}\,\mathrm{Hz}$ & $\lesssim 10^{-18}$ \\
\hhline{|---|}
\end{tabular}
\caption{\justifying Expected GW sources in the (sub-)millihertz frequency range~\cite{Gair2013}: massive black-hole mergers (MBHMs), extreme mass-ratio inspirals (EMRI), and galactic binaries (GBs).}
\label{GW sources}
\endgroup
\end{table}

The anticipated effect can be probed by performing a frequency- and angle-resolving photon number measurement and searching for an excess or deficit of photons with specific wave vectors. To retrieve the maximum information from the measurement, we need a spectral resolution $\Delta k$ much smaller than the characteristic scale at which $\langle \delta n_{\boldsymbol{k}}\rangle$ varies significantly. The latter is given by the peak width of the function $f(\delta_k, T)$, which is $\sim T^{-1}$ (see Fig.~\ref{fig: f}). Since the spectral resolution is limited by the photon collection time $\Delta k\gtrsim T_\text{col}^{-1}$, the ideal resolution requires $T_\text{col}\gg T$. To ideally resolve the angular emission, we have to collect the photons from a small solid angle $\Delta\Omega\ll1$ over which the function $g(\theta,\varphi)$ is approximately constant. Importantly, the ideal resolution is not necessary for detection. In fact, the predicted photon imbalance can be detected as long as the correction does not average out within the detection window $\Delta k\Delta\Omega$. In particular, for a generic evolution time $T$, the correction does not vanish when integrated over all positive (or negative) $\delta_k$. Therefore, it is enough to detect an angle-dependent excess (or deficit) of photons with $\delta_k>0$ with respect to those with $\delta_k<0$ to retrieve some information about the GW.

The magnitude of the GW correction $\langle \delta n_{\boldsymbol{k}}\rangle$ is controlled by $C_{\boldsymbol{k}}\propto\mathcal{A}k/\omega$ (see Eq.~\eqref{emission contributions}), in which the smallness of the GW amplitude $\mathcal{A}$ is (at least partially) balanced by the large factor $k/\omega$. Specifically, for millihertz-frequency GWs and $k$ in the visible range ($\sim10^{14}\,{\rm Hz}$), the latter factor is $\sim10^{17}$. The relevant GW sources in the millihertz frequency range are listed in Table~\ref{GW sources}. Notably, for some of the strongest signals sourced by massive black-hole binaries, we have $C_{\boldsymbol{k}}\gtrsim1$ and the expansion in powers of $C_{\boldsymbol{k}}$ leading to Eq.~\eqref{emission contributions} is not accurate --- one has to take into account other sidebands, which have non-negligible amplitudes. Here we focus on the (more usual) case in which $C_{\boldsymbol{k}}\ll1$, but the effect is still greatly enhanced (as compared to the bare GW amplitude). Note that the same amplification mechanism (originating from the GW-induced phase modulation in Eq.~\eqref{mode solutions}) is used in interferometric detectors~\cite{Maggiore2007}, resulting in observable phase shifts.

\textit{Fisher information---}To investigate how much information about the GW can be extracted from the system, consider the estimation of the GW amplitude $\mathcal{A}$ based on the (ideal-resolution) measurement of the observable $\hat{n}_{\boldsymbol{k}}$. The minimal achievable uncertainty $\delta\mathcal{A}$ of such an estimation is related to the classical Fisher information $\mathcal{I}_\text{C}(\mathcal{A})$ associated with the measurement and the number $M$ of independent repetitions of the measurement through $\delta\mathcal{A}=1/\sqrt{M\mathcal{I}_\text{C}(\mathcal{A})}$. Treating $\langle n_{\boldsymbol{k}}\rangle$ as a probability density for a random variable $\boldsymbol{k}$ conditioned on the value of $\mathcal{A}$, we can calculate the classical Fisher information using the standard formula~\cite{Fisher1922} $\mathcal{I}_\text{C}(\mathcal{A})=\int\mathrm{d}^3\boldsymbol{k}(\partial_{\mathcal{A}}\langle n_{\boldsymbol{k}}\rangle)^2/\langle n_{\boldsymbol{k}}\rangle$. Up to the lowest (zeroth) order in $\mathcal{A}$ this gives (see Appendix~\ref{Appendix: Fisher information})
\begin{equation}\label{classical Fisher information}
    \mathcal{I}_\text{C}(\mathcal{A})=\frac{\bar{n}(T)}{3}\left(\frac{\omega_0}{\omega}\right)^2\cos^2(\omega T/2+\phi_i)[1-\sinc(\omega T)],
\end{equation}
where $\bar{n}(T)\equiv\gamma_0 T$ is the expected total number of photons emitted in time $T$. Note that Eq.~\eqref{classical Fisher information} implicitly assumes shot-noise limited measurements --- we do not include here the technical noise and systematics necessarily present in a real experiment.

The ultimate upper bound on the classical Fisher information is set by the quantum Fisher information $\mathcal{I}_\text{C}(\mathcal{A})\leq\mathcal{I}_\text{Q}(\mathcal{A})$. The latter quantity is measurement-independent and represents the maximum amount of information (about a given parameter) that can be extracted from the quantum state under an optimal measurement strategy~\cite{Braunstein1994}. For the joint atom-field state $\ket{\psi(t_f,t_i)}$, the quantum Fisher information associated with the estimation of $\mathcal{A}$ is given by $\mathcal{I}_\text{Q}(\mathcal{A})=4(\braket{\partial_{\mathcal{A}}\psi(t_f,t_i)}{\partial_{\mathcal{A}}\psi(t_f,t_i)}-|\braket{\psi(t_f,t_i)}{\partial_{\mathcal{A}}\psi(t_f,t_i)}|^2)$ and reads (see Appendix~\ref{Appendix: Fisher information})
\begin{equation}\label{quantum Fisher information}
    \mathcal{I}_\text{Q}(\mathcal{A})=\frac{\bar{n}(T)}{3}\left(\frac{\omega_0}{\omega}\right)^2[1-\cos(\omega T+2\phi_i)\sinc(\omega T)].
\end{equation}
Note that at times $T_m=2(m\pi-\phi_i)/\omega$ with $m\in\mathbb{N}$, the classical and quantum Fisher information coincide, and thus, at those specific times, the photon-number measurement provides maximum information about $\mathcal{A}$. For this reason, we call the times $T_m$ the optimal evolution times.

The optimal evolution times are analogous to the optimal effective arm lengths (effectively determining the GW-light interaction time) in interferometric detectors~\cite{Maggiore2007}. In interferometers, this effective length is practically limited by the physical length of the arms and the quality of the mirrors used to enhance it. In contrast, in the present scheme, the evolution time is ultimately constrained by the lifetime of the atomic transition under consideration. Notably, for the narrow-linewidth transitions used in optical atomic clocks~\cite{Ludlow2015}, such as the $^1S_0 \leftrightarrow {}^3P_0$ transition in $^{87}$Sr with a lifetime of approximately $100$ seconds~\cite{Dorscher2018, Muniz2021, Lu2024, Dolde2025}, the available interaction time is generally much longer than that achievable in ground-based interferometers (typically $\sim 10^{-3}$ seconds for LIGO and Virgo). This suggests that the predicted modifications to the atomic emission spectrum could be advantageous for detecting GWs at lower frequencies, which are beyond the reach of current ground-based interferometers.

\textit{Discussion---}For the GW to be (in principle) detectable through this mechanism, the measurement-specific variance must be $\delta\mathcal{A}\lesssim \mathcal{A}$, which means $\mathcal{A}\gtrsim1/\sqrt{M\mathcal{I}_\text{C}(\mathcal{A})}$. This puts a constraint on the minimal number $M$ of repetitions of the experiment needed for GW detection, namely $M\gtrsim(\mathcal{A}^2\mathcal{I}_\text{C}(\mathcal{A}))^{-1}$. In experiments with many independent atoms, we can think of $M$ as the number of atoms used. Since Eq.~\eqref{classical Fisher information} has been derived using perturbation theory, it gives reliable predictions only for times much smaller than the lifetime of the excited state, $\gamma_0T\ll1$. Therefore, in Appendix~\ref{Appendix: Weisskopf-Wigner} we outline the spontaneous emission description using Weisskopf-Wigner theory~\cite{Agarwal2013}, which is valid for times comparable to and larger than the excited state lifetime, $\gamma_0T\gtrsim1$. In this long-evolution-time limit, the classical Fisher information scales with the square of the quality factor of the transition $Q=\omega_0/\gamma_0$, and in the limit of low GW frequency, $\omega\ll\gamma_0$, it reduces to $\mathcal{I}_\text{C}(\mathcal{A})=(Q^2/3)\cos^2\phi_i$. Thus, the minimal number of atoms required for the detection of such GWs is $M\sim (Q\mathcal{A})^{-2}$. Note that the same scaling can be obtained from Eq.~\eqref{classical Fisher information} by setting the evolution time to $T=1/\gamma_0$ (lifetime of the excited state) and considering the low-frequency limit, $\omega T\ll1$.

The strontium-87 $^1S_0 \leftrightarrow {}^3P_0$ transition used in the most precise optical atomic clocks to date~\cite{Bothwell2022, Zheng2022} has the natural linewidth $\gamma_0/(2\pi)\approx1.35\,{\rm mHz}$~\cite{Muniz2021}, and the quality factor $Q\approx3.18\times10^{17}$. Therefore, to be able to resolve the GW signal in the sub-millihertz frequency range with amplitude $\mathcal{A}\sim10^{-20}-10^{-21}$ (corresponding to the expected sensitivity of LISA~\cite{Robson2019, Babak2021}), one would have to collect the emitted photons from at least $M\sim10^{6}-10^{8}$ atoms. For example, assuming a gravitational wave with amplitude $\mathcal{A}=10^{-21}$ and frequency $\omega/(2\pi)=0.1\,{\rm mHz}$, one would need at least $M\approx3.15\times10^7$ atoms according to our estimates. These numbers would be even lower for the nuclear transition in $^{229}$Th, with a quality factor $Q\sim10^{19}-10^{20}$~\cite{Campbell2012, Tiedau2024}, and GWs with frequencies $\omega\lesssim10^{-5}\,{\rm Hz}$ (the latter being the linewidth of the transition). Note that these are just rough estimates assuming shot-noise-limited measurements. To assess the practical feasibility of GW detection using the scheme proposed in this work, a thorough analysis of additional noise sources is essential. These include standard sources encountered in high-precision fluorescence spectroscopy, such as readout noise, environmental disturbances, and trap-specific noise in the case of confined atoms.

\textit{Conclusion---}We have shown that GWs modulate the quantum field interacting with atoms, leading to distinctive, angle-dependent features in the spontaneous emission spectrum—manifesting as frequency shifts or sidebands. This reveals a new and fundamental imprint of spacetime dynamics within quantum field theory on curved backgrounds, with promising experimental implications. Our results establish a concrete framework for exploring gravitational effects in atom–light interactions, opening new avenues for quantum probes of gravity. Beyond their foundational significance, these findings motivate the development of detection schemes that harness signatures of GWs in atomic spectra, together with detailed studies of realistic noise and sensitivity in future experiments.

\textit{Acknowledgments---}We thank Igor Pikovski, Hendrik Ulbricht, T. Rick Perche, Rafał Demkowicz-Dobrzański, and Dhruva Ganapathy for helpful discussions and useful comments. J.P., N.A., and M.Z. acknowledge the Knut and Alice Wallenberg Foundation through a Wallenberg Academy Fellowship No.\ 2021.0119. S.Q.~is funded in part by the Wallenberg Initiative on Networks and Quantum Information (WINQ) and in part by the Marie Skłodowska--Curie Action IF programme \textit{Nonlinear optomechanics for verification, utility, and sensing} (NOVUS) -- Grant-Number 101027183. Nordita is funded in part by NordForsk. D.B. acknowledges support by the EU EIC Pathfinder project QuCoM (101046973).

\bibliography{biblio}

\onecolumngrid

\appendix

\section{Orthonormal modes}\label{Appendix: orthonormal modes}
Scalar field quantization used in the present work was originally developed in Ref.~\cite{Garriga1991}. For the gravitational wave metric of the form
\begin{equation}\label{app: metric}
    \mathrm{d}s^2=-\mathrm{d}u\mathrm{d}v+g_{ab}(u)\mathrm{d}x^a\mathrm{d}x^b,
\end{equation}
where $u$ and $v$ are the null coordinates and $a,b\in\{1,2\}$, the solutions of the corresponding (massless) Klein-Gordon equation, $\Box\phi(x)=0$, have the following form:
\begin{equation}\label{app: modes}
    U_{k_vk_a}(x)=\frac{(\det g_{ab}(u))^{-1/4}}{\sqrt{(2\pi)^32k_v}}\exp\left[\mathrm{i}k_ax^a-\mathrm{i}k_vv-\frac{\mathrm{i}}{4k_v}\int_0^u\mathrm{d}u g^{ab}k_ak_b\right].
\end{equation}
These are orthonormal with respect to the scalar product
\begin{equation}
    (\psi,\chi)=-\mathrm{i}\int_{u=\text{const.}}\mathrm{d}v\prod_a\mathrm{d}x^a\left(\psi\partial_v\chi^*-\chi^*\partial_v\psi\right),
\end{equation}
defined on a hypersurface $u=\text{const}$, and thus can be used to expand the scalar field operator in the usual way
\begin{equation}\label{app: field expansion}
    \hat{\phi}(x)=\int\mathrm{d}k_v\prod_a\mathrm{d}k_a\left[U_{k_vk_a}(x)\hat{b}_{k_vk_a}+U_{k_vk_a}^*(x)\hat{b}_{k_vk_a}^\dagger\right],
\end{equation}
where $\hat{b}_{k_vk_a}$ and $\hat{b}_{k_vk_a}^\dagger$ are the annihilation and creation operators, respectively, satisfying standard commutation relations $[\hat{b}_{k_vk_a},\hat{b}_{l_vl_a}^\dagger]=\delta(k_v-l_v)\delta^{(2)}(k_a-l_a)$, etc.

Let us rewrite the modes from Eq.~\eqref{app: modes} in the form suitable for this work. We specify $g_{ab}(u)$:
\begin{equation}
    g_{ab}(u)=\begin{pmatrix}
        1+\mathcal{A}\cos (\omega u) & 0 \\
        0 & 1-\mathcal{A}\cos (\omega u)
    \end{pmatrix},
\end{equation}
which leads to
\begin{equation}
    \det g_{ab}(u)=1-\mathcal{A}^2\cos^2(\omega u),\qquad \int_0^u\mathrm{d}u g^{ab}k_ak_b=(k_1^2+k_2^2)u-(k_1^2-k_2^2)\frac{\mathcal{A}}{\omega}\sin(\omega u).
\end{equation}
We introduce coordinates $t=(u+v)/2$, $x=x^1$, $y=x^2$, and $z=(v-u)/2$, in which the metric in Eq.~\eqref{app: metric} is transformed to the one in Eq.~\eqref{metric}. Denoting $k_1\equiv k_x$ and $k_2\equiv k_y$, and introducing $k_v\equiv (k-k_z)/2$ with $k\equiv\sqrt{k_x^2+k_y^2+k_z^2}$, the modes in Eq.~\eqref{app: modes} can be rewritten as follows:
\begin{equation}
    U_{k_vk_a}(x)=\frac{(1-\mathcal{A}^2\cos^2[\omega(t-z)])^{-1/4}}{\sqrt{(2\pi)^32k_v}}\exp\left[-\mathrm{i}(kt-\boldsymbol{k}\cdot\boldsymbol{x})+\mathrm{i}\mathcal{A}\frac{k_x^2-k_y^2}{2\omega(k-k_z)}\sin[\omega(t-z)]\right],
\end{equation}
where the bold symbols $\boldsymbol{k}$ and $\boldsymbol{x}$ correspond to the spatial components of the four-vectors. Now, we want to change the integration variable in Eq.~\eqref{app: field expansion} from $k_v$ to $k_z$, according to
\begin{equation}
    \int_0^\infty\mathrm{d}k_v=-\int_{-\infty}^\infty\mathrm{d}k_z\frac{\partial k_v}{\partial k_z}=\frac{1}{2}\int_{-\infty}^\infty\mathrm{d}k_z\left(1-\frac{\partial k}{\partial k_z}\right)=\int_{-\infty}^\infty\mathrm{d}k_z\frac{k_v}{k}.
\end{equation}
We introduce new modes
\begin{equation}\label{app: new modes}
    u_{\boldsymbol{k}}(x)\equiv\sqrt{\frac{k_v}{k}}U_{k_vk_a}(x)=\frac{(1-\mathcal{A}^2\cos^2[\omega(t-z)])^{-1/4}}{\sqrt{(2\pi)^32k}}\exp\left[-\mathrm{i}(kt-\boldsymbol{k}\cdot\boldsymbol{x})+\mathrm{i}\mathcal{A}\frac{k_x^2-k_y^2}{2\omega(k-k_z)}\sin[\omega(t-z)]\right],
\end{equation}
and a new set of ladder operators $\hat{a}_{\boldsymbol{k}}^{(\dagger)}\equiv\sqrt{\frac{k_v}{k}}\hat{b}_{k_vk_a}^{(\dagger)}$, satisfying $[\hat{a}_{\boldsymbol{k}},\hat{a}_{\boldsymbol{l}}^\dagger]=\delta(k_x-l_x)\delta(k_y-l_y)\delta(k_z-l_z)\equiv\delta^{(3)}(\boldsymbol{k}-\boldsymbol{l})$ [note that $\delta(k_v-l_v)=\frac{k}{k_v}\delta(k_z-l_z)$] with other commutators vanishing, and expand the field operator in terms of those
\begin{equation}
    \hat{\phi}(x)=\int\mathrm{d}^3\boldsymbol{k}\left[u_{\boldsymbol{k}}(x)\hat{a}_{\boldsymbol{k}}+u_{\boldsymbol{k}}^*(x)\hat{a}_{\boldsymbol{k}}^\dagger\right].
\end{equation}
Note that the modes in Eq.~\eqref{app: new modes} are the usual flat-spacetime plane wave modes $\propto \mathrm{e}^{-i(kt-\boldsymbol{k}\cdot\boldsymbol{x})}$ with magnitude and phase modulated by the GW.

To put the modes~\eqref{app: new modes} in a more manageable form, we introduce spherical coordinates in the $\boldsymbol{k}$-space:
\begin{equation}
    \boldsymbol{k}=k(\sin\theta\cos\varphi,\sin\theta\sin\varphi,\cos\theta),
\end{equation}
for which
\begin{equation}
    \frac{k_x^2-k_y^2}{2\omega(k-k_z)}=\frac{k}{\omega}\frac{\sin^2\theta\cos(2\varphi)}{2(1-\cos\theta)}=\frac{k}{\omega}\cos^2(\theta/2)\cos(2\varphi)\equiv\frac{k}{\omega}g(\theta,\varphi),
\end{equation}
where we introduced the function $g(\theta,\varphi)$, used in the main text, for the brevity of notation. At this point, we note that typically the frequency of the relevant quantum field modes will be much greater than the GW frequency, and thus the factor $k/\omega\gg1$. Indeed, in the case considered in this work (optical modes $k\sim10^{14}\;{\rm Hz}$ and GW frequencies in the millihertz range, $\omega\sim10^{-3}\;{\rm Hz}$), we have $k/\omega\sim10^{17}$. Therefore, the amplitude of phase modulation in Eq.~\eqref{app: new modes} will be much larger than the amplitude of magnitude modulation, and the latter effect can be neglected. With this simplification, we can write the modes in the following form
\begin{equation}\label{app: final modes}
    u_{\boldsymbol{k}}(x)=\frac{1}{\sqrt{(2\pi)^32k}}\exp\left[-\mathrm{i}(kt-\boldsymbol{k}\cdot\boldsymbol{x})+\mathrm{i}\mathcal{A}\frac{k}{\omega}g(\theta,\varphi)\sin[\omega(t-z)]\right]\equiv\frac{1}{\sqrt{(2\pi)^32k}}\mathrm{e}^{-\mathrm{i}[kt-\boldsymbol{k}\cdot\boldsymbol{x}-C_{\boldsymbol{k}}\sin(\omega t)]},
\end{equation}
where $C_{\boldsymbol{k}}\equiv\mathcal{A}\frac{k}{\omega}g(\theta,\varphi)$. Note that Eq.~\eqref{app: final modes} describes the usual flat-spacetime plane waves with periodically modulated phase. This phase modulation is the same effect that is detected in interferometric setups~\cite{Maggiore2007}.

\section{System evolution}\label{Appendix: spontaneous emission}
Consider a two-level system (atom) interacting with the scalar field through the interaction Hamiltonian (in the interaction picture)
\begin{equation}\label{app: monopole Hamiltonian}
    \hat{H}_\text{I}(\tau)=\varepsilon\hat{m}(\tau)\hat{\phi}(x(\tau)).
\end{equation}
Here, $\tau$ is the proper time of the atom, $\varepsilon\ll1$ is the small coupling constant, $\hat{m}(\tau)=\mathrm{e}^{\mathrm{i}\omega_0 \tau}\hat{\sigma}_+ + \mathrm{e}^{-\mathrm{i}\omega_0 \tau}\hat{\sigma}_-$ is the monopole operator, and $x(\tau)$ is the trajectory of the atom.

Assume initial atom-field state (at time $\tau_i$) of the form $\ket{\psi_0}=\ket{e,0}$, with the atom in the excited state $\ket{e}$ and field in the vacuum state $\ket{0}$. Up to second order in $\varepsilon$, the state of the system at a later time $\tau_f$ is given by
\begin{equation}\label{app: state perturbative expansion}
    \ket{\psi(\tau_f,\tau_i)}=\ket{\psi_0}-\mathrm{i}\int_{\tau_i}^{\tau_f}\dd \tau\hat{H}_I(\tau)\ket{\psi_0}-\int_{\tau_i}^{\tau_f}\dd \tau\int_{\tau_i}^{\tau}\dd \tau'\hat{H}_I(\tau)\hat{H}_I(\tau')\ket{\psi_0}.
\end{equation}
On the other hand, assuming the rotating-wave approximation (and thus neglecting the amplitudes of processes where the atom and the field are excited or de-excited simultaneously), we conclude that the atom can emit at most one photon, and therefore the state at time $\tau_f$ can be written as
\begin{equation}\label{app: state at any time}
    \ket{\psi(\tau_f,\tau_i)}=\alpha(\tau_f,\tau_i)\ket{e,0}+\int\mathrm{d}^3\boldsymbol{k}\beta_{\boldsymbol{k}}(\tau_f,\tau_i)\ket{g,1_{\boldsymbol{k}}}.
\end{equation}
To get the coefficients $\alpha$ and $\beta_{\boldsymbol{k}}$ we take the inner product of RHS in Eq.~\eqref{app: state perturbative expansion} with $\ket{e,0}$ and $\ket{g,1_{\boldsymbol{k}}}$, respectively. Up to the second order in $\varepsilon$, they are given by
\begin{equation}
    \begin{split}
        &\alpha(\tau_f,\tau_i)=1-\int_{\tau_i}^{\tau_f}\mathrm{d}\tau\int_{\tau_i}^{\tau}\mathrm{d}\tau'\bra{e,0}\hat{H}_I(\tau)\hat{H}_I(\tau')\ket{e,0},\\
        &\beta_{\boldsymbol{k}}(\tau_f,\tau_i)=-\mathrm{i}\int_{\tau_i}^{\tau_f}\mathrm{d}\tau\bra{g,1_{\boldsymbol{k}}}\hat{H}_I(\tau)\ket{e,0}.
    \end{split}
\end{equation}
Inserting the Hamiltonian~\eqref{app: monopole Hamiltonian} and using
\begin{equation}
    \begin{split}
        &\bra{e}\hat{m}(\tau)\hat{m}(\tau')\ket{e}=\mathrm{e}^{\mathrm{i}\omega_0(\tau-\tau')},\qquad \bra{g}\hat{m}(\tau)\ket{e}=\mathrm{e}^{-\mathrm{i}\omega_0\tau},\\
        &\bra{0}\hat{\phi}(x(\tau))\hat{\phi}(x(\tau'))\ket{0}=\int\mathrm{d}^3\boldsymbol{k}u_{\boldsymbol{k}}(x(\tau))u_{\boldsymbol{k}}^*(x(\tau')),\qquad \bra{1_{\boldsymbol{k}}}\hat{\phi}(x(\tau))\ket{0}=u_{\boldsymbol{k}}^*(x(\tau)),
    \end{split}
\end{equation}
we get
\begin{equation}
    \begin{split}
        &\alpha(\tau_f,\tau_i)=1-\varepsilon^2\int_{\tau_i}^{\tau_f}\mathrm{d}\tau\int_{\tau_i}^{\tau}\mathrm{d}\tau'\int\mathrm{d}^3\boldsymbol{k}\mathrm{e}^{\mathrm{i}\omega_0(\tau-\tau')}u_{\boldsymbol{k}}(x(\tau))u_{\boldsymbol{k}}^*(x(\tau')),\\
        &\beta_{\boldsymbol{k}}(\tau_f,\tau_i)=-\mathrm{i}\varepsilon\int_{\tau_i}^{\tau_f}\mathrm{d}\tau\mathrm{e}^{-\mathrm{i}\omega_0\tau}u_{\boldsymbol{k}}^*(x(\tau)).
    \end{split}
\end{equation}
Coefficients $\alpha$ and $\beta_{\boldsymbol{k}}$ are the probability amplitudes for the atom to stay excited, or deexcite emitting a photon with wave-vector $\boldsymbol{k}$, respectively. From now on, we will focus on the latter one.

Let us specify the considerations to the metric~\eqref{metric}, and assume that the atom follows a static trajectory at $x(t)=(t,\boldsymbol{0})$, which is a geodesic with proper time $t$. In this case, the coefficient $\beta_{\boldsymbol{k}}$ is given by
\begin{equation}\label{app: beta 1}
    \beta_{\boldsymbol{k}}(t_f,t_i)=\frac{-\mathrm{i}\varepsilon}{\sqrt{(2\pi)^32k}}\int_{t_i}^{t_f}\mathrm{d}t\mathrm{e}^{\mathrm{i}(k-\omega_0)t}\mathrm{e}^{-\mathrm{i}C_{\boldsymbol{k}}\sin(\omega t)}.
\end{equation}
We introduce $\delta_k\equiv k-\omega_0$, and make use of the Jacobi-Anger expansion:
\begin{equation}
    \mathrm{e}^{-\mathrm{i}C_{\boldsymbol{k}}\sin(\omega t)}=\sum_{n=-\infty}^{\infty} J_n(C_{\boldsymbol{k}})\mathrm{e}^{-\mathrm{i}n\omega t},
\end{equation}
where $J_n(x)$ is the $n$-th Bessel function of the first kind. Inserting this to Eq.~\eqref{app: beta 1}, we get integrals of the form
\begin{equation}
    \int_{t_i}^{t_f}\mathrm{d}t\mathrm{e}^{\mathrm{i}(\delta_k-n\omega)t}=(t_f-t_i)\mathrm{e}^{\mathrm{i}(\delta_k-n\omega)(t_f+t_i)/2}\sinc[(\delta_k-n\omega)(t_f-t_i)/2].
\end{equation}
Introducing the evolution duration $T\equiv t_f-t_i$, and the initial phase of the gravitational wave $\phi_i\equiv\omega t_i$, we can rewrite Eq.~\eqref{app: beta 1} as follows:
\begin{equation}\label{app: amplitude sidebands}
    \beta_{\boldsymbol{k}}(t_f,t_i)=\frac{-\mathrm{i}\varepsilon T\mathrm{e}^{\mathrm{i}\delta_k(t_f+t_i)/2}}{\sqrt{(2\pi)^32k}}\sum_{n=-\infty}^\infty J_n(C_{\boldsymbol{k}})\mathrm{e}^{-\mathrm{i}n(\phi_i+\omega T/2)}\sinc[(\delta_k-n\omega)T/2].
\end{equation}
This should be juxtaposed with the flat-spacetime result
\begin{equation}
    \tilde{\beta}_{\boldsymbol{k}}(t_f,t_i)=\frac{-\mathrm{i}\varepsilon}{\sqrt{(2\pi)^32k}}\int_{t_i}^{t_f}\mathrm{d}t\mathrm{e}^{\mathrm{i}\delta_k t}=\frac{-\mathrm{i}\varepsilon T\mathrm{e}^{\mathrm{i}\delta_k(t_f+t_i)/2}}{\sqrt{(2\pi)^32k}}\sinc[\delta_k T/2].
\end{equation}
We conclude that the GW gives rise to the amplitude sidebands at $\delta_k=n\omega$, with magnitudes $\propto J_n(C_{\boldsymbol{k}})$. Given the amplitude $\beta_{\boldsymbol{k}}$, we can calculate the probability density of emitting a photon with wave-vector $\boldsymbol{k}$:
\begin{equation}\label{app: probability density}
    \begin{split}
        p(\boldsymbol{k})\equiv\left|\beta_{\boldsymbol{k}}(t_f,t_i)\right|^2=&\frac{\varepsilon^2 T^2}{(2\pi)^32k}\sum_{n,m=-\infty}^\infty J_n(C_{\boldsymbol{k}})J_m(C_{\boldsymbol{k}})\mathrm{e}^{-\mathrm{i}(n-m)(\phi_i+\omega T/2)}\sinc[(\delta_k-n\omega)T/2]\sinc[(\delta_k-m\omega)T/2]\\
        \equiv&\frac{\gamma_0T^2}{8\pi^2\omega_0k}\sum_{n,m=-\infty}^\infty J_n(C_{\boldsymbol{k}})J_m(C_{\boldsymbol{k}})\mathrm{e}^{-\mathrm{i}(n-m)(\phi_i+\omega T/2)}\sinc[(\delta_k-n\omega)T/2]\sinc[(\delta_k-m\omega)T/2],
    \end{split}
\end{equation}
where $\gamma_0\equiv\frac{\varepsilon^2\omega_0}{2\pi}$ is the natural linewidth (or the decay rate) of the transition. Note that, since we consider only vacuum and single-photon states of the field, the probability density $p(\boldsymbol{k})$ coincides with the expected number of photons $\langle n_{\boldsymbol{k}}\rangle\equiv\bra{\psi(t_f,t_i)}\hat{a}_{\boldsymbol{k}}^\dagger\hat{a}_{\boldsymbol{k}}\ket{\psi(t_f,t_i)}$ emitted in the time window $t\in[t_i,t_f]$.

Let us now specify to the regime of $C_{\boldsymbol{k}}\ll1$ (meaning $\mathcal{A}\frac{k}{\omega}\ll1$) and expand $\langle n_{\boldsymbol{k}}\rangle$ ($=p(\boldsymbol{k})$) up to the first order in $C_{\boldsymbol{k}}$. The only terms in Eq.~\eqref{app: probability density} that contribute to this order are $\propto J_0^2(C_{\boldsymbol{k}})=1+\mathcal{O}(C_{\boldsymbol{k}}^2)$ and $\propto J_0(C_{\boldsymbol{k}}) J_{\pm 1}(C_{\boldsymbol{k}})=\pm\frac{C_{\boldsymbol{k}}}{2}+\mathcal{O}(C_{\boldsymbol{k}}^3)$. Therefore, up to the first order in $C_{\boldsymbol{k}}$, the expected number of emitted photons is equal to
\begin{equation}
    \begin{split}
        \langle n_{\boldsymbol{k}}\rangle&=\frac{\gamma_0T^2}{8\pi^2\omega_0k}\left[\sinc^2(\delta_k T/2)+C_{\boldsymbol{k}}\cos(\phi_i+\omega T/2)\sinc(\delta_k T/2)\left(\sinc[(\delta_k-\omega) T/2]-\sinc[(\delta_k+\omega) T/2]\right)\right]\\
        &\equiv\langle\tilde{n}_{\boldsymbol{k}}\rangle+\langle\delta n_{\boldsymbol{k}}\rangle,
    \end{split}
\end{equation}
where we split the result into a flat spacetime contribution $\langle\tilde{n}_{\boldsymbol{k}}\rangle$ and a GW correction $\langle\delta n_{\boldsymbol{k}}\rangle$, given by [here we restore $C_{\boldsymbol{k}}=\mathcal{A}\frac{k}{\omega}g(\theta,\varphi)$]
\begin{equation}\label{app: emission contributions}
    \begin{split}
        &\langle \tilde{n}_{\boldsymbol{k}}\rangle=\frac{\gamma_0T^2}{8\pi^2\omega_0k}\sinc^2(\delta_k T/2),\\
        &\langle \delta n_{\boldsymbol{k}}\rangle=\frac{\gamma_0T^2}{8\pi^2\omega_0k}\mathcal{A}\frac{k}{\omega}g(\theta,\varphi)\cos(\phi_i+\omega T/2)\sinc(\delta_k T/2)\left(\sinc{[(\delta_k-\omega)T/2]}-\sinc{[(\delta_k+\omega)T/2]}\right).
    \end{split}
\end{equation}
Note that the GW correction in Eq.~\eqref{app: emission contributions} originates from interference between the carrier (located around $\delta_k=0$) and the first sidebands ($\delta_k=\pm\omega$). Introducing the function
\begin{equation}
    f(\delta_k,T)=\sinc(\delta_k T/2)\left(\sinc{[(\delta_k-\omega)T/2]}-\sinc{[(\delta_k+\omega)T/2]}\right),
\end{equation}
we can write the GW correction in a more compact form
\begin{equation}
    \langle \delta n_{\boldsymbol{k}}\rangle=\frac{\gamma_0T^2}{8\pi^2\omega_0k}\mathcal{A}\frac{k}{\omega}g(\theta,\varphi)f(\delta_k,t)\cos(\phi_i+\omega T/2).
\end{equation}

Notably, the GW correction vanishes when integrated over $\boldsymbol{k}$ (indeed, this is ensured by the factor $g(\theta,\varphi)=\cos^2(\theta/2)\cos(2\varphi)$, which vanishes when integrated over $\varphi$). Therefore, the total number of emitted photons $\Bar{n}(T)$ is unaffected by the GW up to the first order in GW amplitude, and reads
\begin{equation}\label{app: total number of emitted photons 0}
    \Bar{n}(T)=\int\mathrm{d}^3\boldsymbol{k}\langle\tilde{n}_{\boldsymbol{k}}\rangle=\int_0^{2\pi}\mathrm{d}\varphi\int_0^\pi\mathrm{d}\theta\sin\theta\int_0^\infty\mathrm{d}k k^2\frac{\gamma_0T^2}{8\pi^2\omega_0k}\sinc^2(\delta_k T/2)=\frac{\gamma_0T^2}{2\pi\omega_0}\int_0^\infty\mathrm{d}kk\sinc^2(\delta_k T/2).
\end{equation}
The integral over $\mathrm{d}k$ is formally infinite and has to be regularized by introducing a high-frequency cut-off~\cite{Bethe1947}. The $\sinc^2(\delta_k T/2)$ is sharply peaked around $\delta_k=0$ (corresponding to $k=\omega_0$), therefore, the factor $k$ within the integral can be approximated by $\omega_0$ (indeed, in the limit of $T\gg\omega_0^{-1}$, considered here, $\sinc^2(\delta_kT/2)$ behaves as a Dirac delta~\cite{Loudon2000, Gerry2005}). After this simplification, we can extend the limits of integration to $\pm\infty$, which leads to
\begin{equation}\label{app: total number of emitted photons 1}
    \bar{n}(T)=\frac{\gamma_0 T^2}{2\pi}\int_{-\infty}^\infty\dd k\sinc^2(\delta_k T/2)=\gamma_0T,
\end{equation}
as expected in the short evolution time limit, $\gamma_0T\ll1$ (in which the perturbative expansion in $\varepsilon$ is valid).

\section{Low frequency regime}\label{Appendix: low frequency regime}
Let us analyze the results from Appendices~\ref{Appendix: orthonormal modes} and~\ref{Appendix: spontaneous emission} in the regime of very low GW frequencies such that $\omega T\ll1$. First, we evaluate the modes from Eq.~\eqref{app: final modes} at $\boldsymbol{x}=\boldsymbol{0}$ and $t=t_i+\Delta t$ (with $\Delta t\in[0,T]$)
\begin{equation}
    \begin{split}
        u_{\boldsymbol{k}}(t)=&\exp\left(\mathrm{i}\mathcal{A}\frac{k}{\omega}g(\theta,\varphi)\sin{[\omega (t_i+\Delta t)]}\right)\frac{\mathrm{e}^{-\mathrm{i}k(t_i+\Delta t)}}{\sqrt{(2\pi)^32k}}\\
        =&\exp\left(\mathrm{i}\mathcal{A}\frac{k}{\omega}g(\theta,\varphi)\left[\sin(\omega t_i)\cos(\omega\Delta t)+\sin(\omega \Delta t)\cos(\omega t_i)\right]\right)\frac{\mathrm{e}^{-\mathrm{i}k(t_i+\Delta t)}}{\sqrt{(2\pi)^32k}}.
    \end{split}
\end{equation}
Introducing $\phi_i\equiv\omega t_i$ and $\tilde{\phi}_i\equiv -kt_i$ and expanding the argument of the first exponential function up to the first order in $\omega\Delta t$, we arrive at
\begin{equation}\label{app: modes in the low frequency limit}
    u_{\boldsymbol{k}}(t)=\exp\left[\mathrm{i}(\tilde{\phi}_i+\mathcal{A}\frac{k}{\omega}g(\theta,\varphi)\sin\phi_i)\right]\frac{\exp\left[-\mathrm{i}k\left(1-\mathcal{A}g(\theta,\varphi)\cos\phi_i\right)\Delta t\right]}{\sqrt{(2\pi)^32k}}.
\end{equation}
The first exponential factor is just a constant ($\Delta t$-independent) phase, and does not affect $\langle n_{\boldsymbol{k}}\rangle$. On the other hand, the second exponential factor oscillates with $\Delta t$ with frequency $k\left[1-\mathcal{A}g(\theta,\varphi)\cos\phi_i\right]$. Therefore, GW in the low-frequency limit gives rise to an angle-dependent frequency shift of the field modes
\begin{equation}\label{app: frequency shift}
    k\to k\left[1-\mathcal{A}g(\theta,\varphi)\cos\phi_i\right].
\end{equation}

Now, let us analyze the behavior of the GW correction $\langle\delta n_{\boldsymbol{k}}\rangle$ in the low-frequency regime, $\omega T\ll1$. We start by expanding Eq.~\eqref{app: emission contributions} up to the first order in $\omega T$. We use
\begin{equation}
    \sinc[(\delta_k\mp\omega)T/2]=\sinc(\delta_k T/2)\mp\sinc'(\delta_k T/2)\omega T/2+\mathcal{O}\left((\omega T)^2\right),
\end{equation}
which leads to
\begin{equation}\label{app: correction in the low frequency limit}
    \begin{split}
        \langle\delta n_{\boldsymbol{k}}\rangle=&-\frac{\gamma_0T^2}{8\pi^2\omega_0k}\mathcal{A}g(\theta,\varphi)\cos\phi_i kT\sinc(\delta_kT/2)\sinc'(\delta_kT/2)\\
        =&-\frac{\gamma_0T^2}{8\pi^2\omega_0k}\left[\frac{\mathrm{d}}{\mathrm{d}\delta_k}\sinc^2(\delta_k T/2)\right]k\mathcal{A}g(\theta,\varphi)\cos\phi_i.
    \end{split}
\end{equation}
Combining this with the flat spacetime contribution $\langle\tilde{n}_{\boldsymbol{k}}\rangle$, we get
\begin{equation}\label{app: carrier shift}
    \begin{split}
        \langle n_{\boldsymbol{k}}\rangle=&\frac{\gamma_0T^2}{8\pi^2\omega_0k}\left(\sinc^2(\delta_k T/2)-\left[\frac{\mathrm{d}}{\mathrm{d}\delta_k}\sinc^2(\delta_k T/2)\right]k\mathcal{A}g(\theta,\varphi)\cos\phi_i \right)\\
        \approx&\frac{\gamma_0T^2}{8\pi^2\omega_0k}\sinc^2\left[\left(\delta_k-k\mathcal{A}g(\theta,\varphi)\cos\phi_i\right)\right]\\
        =&\frac{\gamma_0T^2}{8\pi^2\omega_0k}\sinc^2\left[\left(k\left[1-\mathcal{A}g(\theta,\varphi)\cos\phi_i\right]-\omega_0\right)T/2\right],
    \end{split}
\end{equation}
consistent with the frequency shift~\eqref{app: frequency shift}.

Since the correction in Eq.~\eqref{app: correction in the low frequency limit} is independent of the GW frequency $\omega$ (apart from the dependence on the initial phase $\phi_i=\omega t_i$), we cannot retrieve any information about the latter from measurements performed at a particular time instance $t_i$. However, it is still possible to infer $\omega$ by performing a series of measurements at different times $t_i$, and observing how the magnitude of the effect (which is $\propto\cos\phi_i$) changes in time.

\section{Fisher information}\label{Appendix: Fisher information}
\subsection{Classical Fisher information}
Consider the estimation of the GW amplitude $\mathcal{A}$ based on the measured value of $\langle n_{\boldsymbol{k}}\rangle$ for each $\boldsymbol{k}$. Recall that the latter can be treated as a probability density function (see Eq.~\eqref{app: probability density}) that depends on the parameter $\mathcal{A}$, i.e., $\langle n_{\boldsymbol{k}}\rangle\equiv p(\boldsymbol{k};\mathcal{A})$. Note that $p(\boldsymbol{k};\mathcal{A})$ does not integrate to 1, but rather to the total probability of emitting a photon (equal to $\Bar{n}(T)\leq1$), which is independent of $\mathcal{A}$ up to first order (see discussion around Eqs.~\eqref{app: total number of emitted photons 0} and~\eqref{app: total number of emitted photons 1}).

The minimal uncertainty of such an estimation $\delta\mathcal{A}$ is given by the Cram{\'e}r-Rao bound~\cite{Cramer1946, Rao1945}
\begin{equation}
    \delta\mathcal{A}=\frac{1}{\sqrt{M\mathcal{I}_\text{C}(\mathcal{A})}},
\end{equation}
where $M$ is the number of independent repetitions of the measurement, and $\mathcal{I}_\text{C}(\mathcal{A})$ is the classical Fisher information, defined as follows:
\begin{equation}
    \mathcal{I}_\text{C}(\mathcal{A})=\int\mathrm{d}^3\boldsymbol{k}\frac{\left[\partial_{\mathcal{A}}p(\boldsymbol{k};\mathcal{A})\right]^2}{p(\boldsymbol{k};\mathcal{A})}\equiv\int\mathrm{d}^3\boldsymbol{k}\frac{\left(\partial_{\mathcal{A}}\langle n_{\boldsymbol{k}}\rangle\right)^2}{\langle n_{\boldsymbol{k}}\rangle}.
\end{equation}
To get the lowest order term in $\mathcal{A}$, we use Eq.~\eqref{app: emission contributions} and replace $\langle n_{\boldsymbol{k}}\rangle$ in the numerator by $\langle\delta n_{\boldsymbol{k}}\rangle$ (since $\langle\tilde{n}_{\boldsymbol{k}}\rangle$ is independent of $\mathcal{A}$) and the one in the denominator by $\langle\tilde{n}_{\boldsymbol{k}}\rangle$. In this way, we arrive at
\begin{equation}\label{app: classical Fisher information}
    \begin{split}
        \mathcal{I}_\text{C}(\mathcal{A})=&\int\mathrm{d}^3\boldsymbol{k}\frac{\left(\partial_{\mathcal{A}}\langle \delta n_{\boldsymbol{k}}\rangle\right)^2}{\langle \tilde{n}_{\boldsymbol{k}}\rangle}=\int\mathrm{d}^3\boldsymbol{k}\frac{\gamma_0T^2}{8\pi^2\omega_0k}\left(\frac{k}{\omega}\cos(\phi_i+\omega T/2)g(\theta,\varphi)\frac{f(\delta_k,T)}{\sinc(\delta_k T/2)}\right)^2\\
        =&\frac{\gamma_0T^2}{8\pi^2\omega_0\omega^2}\cos^2(\omega T/2+\phi_i)\int_0^{2\pi}\mathrm{d}\varphi\cos^2(2\varphi)\int_0^\pi\mathrm{d}\theta\sin\theta\cos^2(\theta/2)\times\\
        &\times\int_0^\infty\mathrm{d}k k^3(\sinc[(\delta_k-\omega)T/2]-\sinc[(\delta_k+\omega)T/2])^2.
    \end{split}
\end{equation}
The angular integrals are straightforward and give
\begin{equation}\label{app: angular integral}
    \int_0^{2\pi}\dd\varphi\cos^2(2\varphi)=\pi,\qquad\int_0^\pi\dd\theta\sin\theta\cos^4(\theta/2)=\frac{2}{3}.
\end{equation}
As in Eq.~\eqref{app: total number of emitted photons 0}, the integral over $\mathrm{d}k$ is formally divergent and will be handled analogously. We first regularize it by introducing a high-frequency cut-off, and then replace the factor $k^3$ by $\omega_0^3$, noting that the integrand is sharply peaked around that value (more precisely, it is peaked around $\omega_0\pm\omega$, but we will approximate it by $\omega_0$, since $\omega\ll\omega_0$). Then, the integration limits can be extended to $\pm\infty$, and the integral can be performed analytically, giving
\begin{equation}\label{app: frequency integral}
    \begin{split}
        \int_0^\infty\dd k k^3\left(\sinc{[(\delta_k-\omega)T/2]}-\sinc{[(\delta_k+\omega)T/2]}\right)^2\approx&\omega_0^3\int_{-\infty}^\infty\dd k\left(\sinc{[(\delta_k-\omega)T/2]}-\sinc{[(\delta_k+\omega)T/2]}\right)^2\\
        =&\frac{4\pi}{T}[1-\sinc(\omega T)].
    \end{split}
\end{equation}
Therefore, the classical Fisher information is finally given by
\begin{equation}
    \mathcal{I}_\text{C}(\mathcal{A})=\frac{\gamma_0T}{3}\frac{\omega_0^2}{\omega^2}\cos^2(\omega T/2+\phi_i)[1-\sinc(\omega T)]\equiv\frac{\Bar{n}(T)}{3}\frac{\omega_0^2}{\omega^2}\cos^2(\omega T/2+\phi_i)[1-\sinc(\omega T)],
\end{equation}
where $\Bar{n}(T)\equiv\gamma_0T$ is the total emitted number of photons from Eq.~\eqref{app: total number of emitted photons 1}.

\subsection{Quantum Fisher information}\label{Appendix: quantum Fisher information}
The quantum Fisher information $\mathcal{I}_\text{Q}(\mathcal{A})$ gives the absolute upper bound on the information that can be extracted from the system. For an $\mathcal{A}$-dependent state of the quantum system $\ket{\psi(t_f,t_i)}$, the quantum Fisher information is given by
\begin{equation}
    \mathcal{I}_\text{Q}(\mathcal{A})=4(\braket{\partial_{\mathcal{A}}\psi(t_f,t_i)}{\partial_{\mathcal{A}}\psi(t_f,t_i)}-|\braket{\psi(t_f,t_i)}{\partial_{\mathcal{A}}\psi(t_f,t_i)}|^2).
\end{equation}
For the state~\eqref{app: state perturbative expansion}, and up to the lowest order in $\mathcal{A}$ and $\varepsilon$, it reduces to
\begin{equation}\label{app: quantum Fisher information 1}
    \begin{split}
        \mathcal{I}_\text{Q}(\mathcal{A})=&4\braket{\partial_{\mathcal{A}}\psi(t_f,t_i)}{\partial_{\mathcal{A}}\psi(t_f,t_i)}=4\int_{t_i}^{t_f}\dd t\int_{t_i}^{t_f}\dd t'\bra{\psi_0}\left(\partial_{\mathcal{A}}\hat{H}_\text{I}(t)\right)\left(\partial_{\mathcal{A}}\hat{H}_\text{I}(t')\right)\ket{\psi_0}\\
        =&4\varepsilon^2\int_{t_i}^{t_f}\dd t\int_{t_i}^{t_f}\dd t'\bra{e}\hat{m}(t)\hat{m}(t')\ket{e}\bra{0}\left(\partial_{\mathcal{A}}\hat{\phi}(t)\right)\left(\partial_{\mathcal{A}}\hat{\phi}(t')\right)\ket{0}\\
        =&4\varepsilon^2\int\dd^3\boldsymbol{k}\int_{t_i}^{t_f}\dd t\int_{t_i}^{t_f}\dd t'\mathrm{e}^{\mathrm{i}\omega_0(t-t')}\partial_{\mathcal{A}}u_{\boldsymbol{k}}(t)\partial_{\mathcal{A}}u_{\boldsymbol{k}}^*(t')\\
        =&4\varepsilon^2\int\dd^3\boldsymbol{k}\left|\int_{t_i}^{t_f}\dd t\mathrm{e}^{\mathrm{i}\omega_0t}\partial_{\mathcal{A}}u_{\boldsymbol{k}}(t)\right|^2.
    \end{split}
\end{equation}
Using the modes~\eqref{app: final modes} expanded up to first order in $\mathcal{A}$, we get
\begin{equation}\label{app: quantum Fisher information 2}
    \mathcal{I}_\text{Q}(\mathcal{A})=\frac{\gamma_0}{2\pi^2\omega_0\omega^2}\int\dd^3\boldsymbol{k}kg^2(\theta,\varphi)\left|\int_{t_i}^{t_f}\dd t\mathrm{e}^{-\mathrm{i}\delta_k t}\sin{(\omega t)}\right|^2.
\end{equation}
The integral over $\mathrm{d}t$ equals
\begin{equation}
    \int_{t_i}^{t_f}\dd t\mathrm{e}^{-\mathrm{i}\delta_k t}\sin{(\omega t)}=\frac{T\mathrm{e}^{-\mathrm{i}\delta_k T/2}}{2\mathrm{i}}\left(\mathrm{e}^{\mathrm{i}\omega (T/2+t_i)}\sinc[(\delta_k-\omega)T/2]-\mathrm{e}^{-\mathrm{i}\omega (T/2+t_i)}\sinc[(\delta_k+\omega)T/2]\right),
\end{equation}
which, when inserted into Eq.~\eqref{app: quantum Fisher information 2}, gives
\begin{equation}
    \begin{split}
        \mathcal{I}_\text{Q}(\mathcal{A})=\frac{\gamma_0T^2}{8\pi^2\omega_0\omega^2}\int\dd^3\boldsymbol{k}kg^2(\theta,\varphi)\big(&\sinc^2[(\delta_k-\omega)T/2]+\sinc^2[(\delta_k+\omega)T/2]\\
        &-2\cos(\omega T+2\phi_i)\sinc[(\delta_k-\omega)T/2]\sinc[(\delta_k+\omega)T/2]\big).
    \end{split}
\end{equation}
The integral over $\dd^3\boldsymbol{k}$ can be performed analogously to the one in~\eqref{app: classical Fisher information}, which leads to
\begin{equation}
    \mathcal{I}_\text{Q}(\mathcal{A})=\frac{\gamma_0T}{3}\frac{\omega_0^2}{\omega^2}[1-\cos(\omega T+2\phi_i)\sinc(\omega T)]=\frac{\bar{n}(T)}{3}\frac{\omega_0^2}{\omega^2}[1-\cos(\omega T+2\phi_i)\sinc(\omega T)].
\end{equation}

\section{Weisskopf-Wigner approach}\label{Appendix: Weisskopf-Wigner}
The perturbative approach used in Appendix~\ref{Appendix: spontaneous emission} is valid only for evolution times much shorter than the lifetime of the atom. Here, we follow the Weisskopf-Wigner description of spontaneous decay~\cite{Agarwal2013}, which is accurate for times comparable to and longer than the lifetime. As in Appendix~\ref{Appendix: spontaneous emission}, we assume that the state of the system is at any time given by Eq.~\eqref{app: state at any time}. However, this time, to derive the coefficients $\alpha$ and $\beta_{\boldsymbol{k}}$ we use the Schr{\"o}dinger equation with Hamiltonian~\eqref{app: monopole Hamiltonian}. This leads to the following set of equations (here we restrict the considerations to the GW background and atom following the static trajectory $x(t)=(t,\boldsymbol{0})$):
\begin{equation}\label{app: alpha and beta equations}
    \begin{split}
        &\frac{\mathrm{d}}{\mathrm{d}t}\alpha(t,t_i)=-\mathrm{i}\varepsilon\int\mathrm{d}^3\boldsymbol{k}\mathrm{e}^{\mathrm{i}\omega_0 t}u_{\boldsymbol{k}}(t)\beta_{\boldsymbol{k}}(t,t_i),\\
        &\frac{\mathrm{d}}{\mathrm{d}t}\beta_{\boldsymbol{k}}(t,t_i)=-\mathrm{i}\varepsilon\mathrm{e}^{-\mathrm{i}\omega_0 t}u_{\boldsymbol{k}}^*(t)\alpha(t,t_i).
    \end{split}
\end{equation}
We integrate the second equation in~\eqref{app: alpha and beta equations} over time
\begin{equation}\label{app: beta equation}
    \beta_{\boldsymbol{k}}(t,t_i)=-\mathrm{i}\varepsilon\int_{t_i}^t\mathrm{d}t'\mathrm{e}^{-\mathrm{i}\omega_0 t'}u_{\boldsymbol{k}}^*(t')\alpha(t',t_i),
\end{equation}
and insert it into the first equation in~\eqref{app: alpha and beta equations} to get
\begin{equation}\label{app: alpha equation}
    \frac{\mathrm{d}}{\mathrm{d}t}\alpha(t,t_i)=-\varepsilon^2\int\mathrm{d}^3\boldsymbol{k}\int_{t_i}^t\mathrm{d}t'\mathrm{e}^{\mathrm{i}\omega_0(t-t')}u_{\boldsymbol{k}}(t)u_{\boldsymbol{k}}^*(t')\alpha(t',t_i).
\end{equation}
Using the modes~\eqref{app: final modes} and expanding up to the first order in $C_{\boldsymbol{k}}$ (or, equivalently, in $\mathcal{A}$) we find that the equation for $\alpha$ is unaffected by the GW --- the correction $\propto C_{\boldsymbol{k}}$ vanishes when integrated over $\mathrm{d}^3\boldsymbol{k}$ (this follows from the fact that $C_{\boldsymbol{k}}\propto\cos(2\varphi)$, which vanishes when integrated over $\mathrm{d}\varphi$). Therefore, Eq.~\eqref{app: alpha equation} reduces to
\begin{equation}
    \frac{\mathrm{d}}{\mathrm{d}t}\alpha(t,t_i)=-\frac{\varepsilon^2}{2(2\pi)^3}\int\frac{\mathrm{d}^3\boldsymbol{k}}{k}\int_{t_i}^t\mathrm{d}t'\mathrm{e}^{-\mathrm{i}(k-\omega_0)(t-t')}\alpha(t',t_i)=-\frac{\varepsilon^2}{4\pi^2}\int_0^\infty\mathrm{d}kk\int_{t_i}^t\mathrm{d}t'\mathrm{e}^{-\mathrm{i}(k-\omega_0)(t-t')}\alpha(t',t_i).
\end{equation}

Following the standard procedure, we assume that $\alpha(t,t_i)$ varies on a scale $\gamma_0^{-1}\gg\omega_0^{-1}$ (here $\gamma_0$ is the atomic decay rate), and therefore replace $\alpha(t',t_i)$ in the integrand by $\alpha(t,t_i)$. This constitutes the Markov approximation, meaning that the system has no memory of its past. Next, we change the integration variable $t'\to t''=t'-t_i$, so that
\begin{equation}
    \frac{\mathrm{d}}{\mathrm{d}t}\alpha(t,t_i)=-\frac{\varepsilon^2\alpha(t,t_i)}{4\pi^2}\int_0^\infty\mathrm{d}kk\int_{0}^{t-t_i}\mathrm{d}t''\mathrm{e}^{-\mathrm{i}(k-\omega_0)((t-t_i)-t'')},
\end{equation}
and extend the upper limit of the $t''$ integral to $\infty$ (we are interested in times $t-t_i\gg\omega_0^{-1}$). With this simplification, we get
\begin{equation}
    \begin{split}
        \frac{\mathrm{d}}{\mathrm{d}t}\alpha(t,t_i)=&-\frac{\varepsilon^2\alpha(t,t_i)}{4\pi^2}\int_0^\infty\mathrm{d}kk\int_{0}^{\infty}\mathrm{d}t''\mathrm{e}^{-\mathrm{i}(k-\omega_0)((t-t_i)-t'')}\\
        =&-\frac{\varepsilon^2\alpha(t,t_i)}{4\pi^2}\int_0^\infty\mathrm{d}kk\left[\pi\delta(\omega_0-k)+\mathrm{i}\mathcal{P}\left(\frac{1}{\omega_0-k}\right)\right].
    \end{split}
\end{equation}
The real part of this expression gives the decay rate, while the imaginary part --- the Lamb shift of the internal energy. The latter can be absorbed into a redefinition of $\omega_0$ and will be omitted in subsequent calculations. In this way, we obtain
\begin{equation}
    \frac{\mathrm{d}}{\mathrm{d}t}\alpha(t,t_i)=-\frac{\varepsilon^2\alpha(t,t_i)}{4\pi}\int_0^\infty\mathrm{d}kk\delta(\omega_0-k)=-\frac{\gamma_0}{2}\alpha(t,t_i),
\end{equation}
Given the initial condition $\alpha(t_i,t_i)=1$, we find that
\begin{equation}
    \alpha(t,t_i)=\mathrm{e}^{-\gamma_0(t-t_i)/2}.
\end{equation}
Inserting this result back into Eq.~\eqref{app: beta equation}, and expanding it up to the first order in $C_{\boldsymbol{k}}$, we arrive at
\begin{equation}\label{app: beta coefficients}
    \beta_{\boldsymbol{k}}(t,t_i)=\frac{\varepsilon\mathrm{e}^{\mathrm{i}\delta_k t_i}}{\sqrt{(2\pi)^32k}}\left[\frac{1-\mathrm{e}^{-(\gamma_0/2-\mathrm{i}\delta_k)(t-t_i)}}{\mathrm{i}\gamma_0/2+\delta_k}-\frac{C_{\boldsymbol{k}}}{2}\sum_{\sigma=\pm1}\sigma\mathrm{e}^{\mathrm{i}\sigma\phi_i}\frac{1-\mathrm{e}^{-[\gamma_0/2-\mathrm{i}(\delta_k+\sigma\omega)](t-t_i)}}{\mathrm{i}\gamma_0/2+(\delta_k+\sigma\omega)}\right].
\end{equation}
This equation is analogous to Eq.~\eqref{app: amplitude sidebands} --- it predicts the existence of a carrier centered at $\delta_k=0$ and two sidebands at $\delta_k=\pm\omega$ (there are only two sidebands because we expanded up to the first order in $C_{\boldsymbol{k}}$; other sidebands contribute only in higher orders). In the limit of $t-t_i\gg\gamma_0^{-1}$, Eq.~\eqref{app: beta coefficients} reduces to
\begin{equation}
    \beta_{\boldsymbol{k}}(\infty)=\frac{\varepsilon\mathrm{e}^{\mathrm{i}\delta_k t_i}}{\sqrt{(2\pi)^32k}}\left[\frac{1}{\mathrm{i}\gamma_0/2+\delta_k}-\frac{C_{\boldsymbol{k}}}{2}\sum_{\sigma=\pm1}\frac{\sigma\mathrm{e}^{\mathrm{i}\sigma\phi_i}}{\mathrm{i}\gamma_0/2+(\delta_k+\sigma\omega)}\right],
\end{equation}
and leads to the probability density
\begin{equation}
    |\beta_{\boldsymbol{k}}(\infty)|^2=\frac{\gamma_0}{8\pi^2\omega_0k}\left[\frac{1}{\left(\frac{\gamma_0}{2}\right)^2+\delta_k^2}-C_{\boldsymbol{k}}\sum_{\sigma=\pm1}\frac{\left(\delta_k(\delta_k+\sigma\omega)+\left(\frac{\gamma_0}{2}\right)^2\right)\cos\phi_i-\frac{\gamma_0}{2}\omega\sin\phi_i}{\left(\delta_k^2+\left(\frac{\gamma_0}{2}\right)^2\right)\left((\delta_k+\sigma\omega)^2+\left(\frac{\gamma_0}{2}\right)^2\right)}+\mathcal{O}(C_{\boldsymbol{k}}^2)\right].
\end{equation}

We denote $p(\boldsymbol{k};\mathcal{A})\equiv|\beta_{\boldsymbol{k}}(\infty)|^2$ and aim to calculate the classical Fisher information associated with $\mathcal{A}$ estimation as in Appendix~\ref{Appendix: Fisher information}. Note that $p(\boldsymbol{k};\mathcal{A})$ depends on $\mathcal{A}$ only through the $C_{\boldsymbol{k}}$ factor, which satisfies $\partial_{\mathcal{A}}C_{\boldsymbol{k}}=\frac{k}{\omega}g(\theta,\varphi)$. Therefore, up to the lowest (zeroth) order in $\mathcal{A}$, the Fisher information is given by
\begin{equation}
    \mathcal{I}_{\text{C}}(\mathcal{A})=\int\mathrm{d}^3\boldsymbol{k}\frac{\gamma_0}{8\pi^2\omega_0k}\left(\frac{k}{\omega}g(\theta,\varphi)\right)^2\left(\sum_{\sigma=\pm1}\frac{\left(\delta_k(\delta_k+\sigma\omega)+\left(\frac{\gamma_0}{2}\right)^2\right)\cos\phi_i-\frac{\gamma_0}{2}\omega\sin\phi_i}{\left((\delta_k+\sigma\omega)^2+\left(\frac{\gamma_0}{2}\right)^2\right)}\right)^2\left(\left(\frac{\gamma_0}{2}\right)^2+\delta_k^2\right)^{-1}.
\end{equation}
Calculating the integral is completely analogous to the one in Eq.~\eqref{app: classical Fisher information}. The angular integrals are the same as in Eq.~\eqref{app: angular integral} and give a factor of $2\pi/3$. Then, we are left with
\begin{equation}
    \mathcal{I}_{\text{C}}(\mathcal{A})=\frac{\gamma_0}{12\pi\omega_0\omega^2}\int_0^\infty\mathrm{d}kk^3\left(\sum_{\sigma=\pm1}\frac{\left(\delta_k(\delta_k+\sigma\omega)+\left(\frac{\gamma_0}{2}\right)^2\right)\cos\phi_i-\frac{\gamma_0}{2}\omega\sin\phi_i}{\left((\delta_k+\sigma\omega)^2+\left(\frac{\gamma_0}{2}\right)^2\right)}\right)^2\left(\left(\frac{\gamma_0}{2}\right)^2+\delta_k^2\right)^{-1}.
\end{equation}
Following the same reasoning as in Eqs~\eqref{app: total number of emitted photons 0}, and~\eqref{app: frequency integral}, we approximate the factor $k^3\approx\omega_0^3$, and extend the lower limit of the integral to $-\infty$, to get
\begin{equation}
    \begin{split}
        \mathcal{I}_{\text{C}}(\mathcal{A})=&\frac{\gamma_0}{12}\frac{\omega_0^2}{\omega^2}\left[\frac{2}{\gamma_0}-\frac{4\gamma_0\left(\gamma_0\sin\phi_i+\omega\cos\phi_i\right)^2}{(\gamma_0^2+\omega^2)^2}-\frac{\left(\frac{\gamma_0}{2}\cos(2\phi_i)-\omega\sin(2\phi_i)\right)}{\left(\frac{\gamma_0}{2}\right)^2+\omega^2}\right]\\
        \equiv&\frac{1}{12}\frac{Q^2}{\xi^2}\left[2-\frac{4\left(\sin\phi_i+\xi\cos\phi_i\right)^2}{(1+\xi^2)^2}-\frac{\left(\frac{1}{2}\cos(2\phi_i)-\xi\sin(2\phi_i)\right)}{\frac{1}{4}+\xi^2}\right],
    \end{split}
\end{equation}
where in the last step we introduced dimensionless parameters $Q\equiv\frac{\omega_0}{\gamma_0}$ (the quality factor of the atomic transition) and $\xi\equiv\frac{\omega}{\gamma_0}$. In the limit of low GW frequency, $\omega\ll\gamma_0$, the classical Fisher information reduces to
\begin{equation}
    \mathcal{I}_{\text{C}}(\mathcal{A})=\frac{Q^2}{3}\cos^2\phi_i.
\end{equation}

\end{document}